\documentclass[aps,prd,twocolumn,groupedaddress,showpacs]{revtex4}

\usepackage{graphics}

\newcommand{\omm}{\Omega_{m0}}

\newcommand{\omq}{\Omega_{Q0}}

\newcommand{\bc}{\begin{center}}
\newcommand{\ec}{\end{center}}
\newcommand{\rat}{\mathcal R_0}
\newcommand{\dub}{w_Q}
\newcommand{\phidot}{\dot{\phi}}
\newcommand{\rhoq}{\rho_Q}
\newcommand{\rhob}{\rho_B}
\newcommand{\be}{\begin{equation}}
\newcommand{\ee}{\end{equation}}
\newcommand{\bitem}{\begin{itemize} \setlength{\itemsep 3pt} }
  \newcommand{\eitem}{\end{itemize}}
\newcommand{\benum}{\begin{enumerate} \setlength{\itemsep 3pt} }
  \newcommand{\eenum}{\end{enumerate}}

\begin{document}

\title{Quintessence Cosmology and the Cosmic Coincidence}
\author{S. A. Bludman}\email[]{bludman@mail.desy.de}
\affiliation{Deutsches Elektronen-Synchrotron DESY, Hamburg\\ University of Pennsylvania, Philadelphia}

\author{M. Roos}\email[]{Matts.Roos@helsinki.fi}
\affiliation{Division of High Energy Physics, University of Helsinki, Helsinki}

\date{\today}

\begin{abstract}
  Within present constraints on the observed smooth energy and its
  equation of state parameter $\dub=P/\rhoq$,
  it is important to find out whether the smooth energy is static
  (cosmological constant) or dynamic (quintessence).
  The most dynamical
  quintessence fields observationally allowed are now still
  fast-rolling and no longer satisfy the tracker approximation if
  the equation of
  state parameter varies moderately with cosmic
  scale $a=1/1+z$.  We are optimistic about distinguishing between a
  cosmological constant and appreciably dynamic quintessence, by
  measuring
  average values for the effective equation of state parameter $\dub(a)$.
  However, reconstructing the
  quintessence potential from observations of any scale dependence
  $\dub(a)$ appears problematic in the near future.  For our flat
  universe, at present dominated by smooth energy in the form of either a cosmological
  constant (LCDM) or quintessence (QCDM), we calculate the asymptotic
  collapsed mass fraction to be maximal at the observed smooth energy/matter
  ratio $\rat \sim 2$.  Identifying this collapsed fraction as a
  conditional probability for habitable galaxies, we infer that the
  prior distribution is flat in $\rat$ or $\omm$.  Interpreting this
  prior as a distribution over {\em theories}, rather than as a
  distribution over unobservable {\em subuniverses}, leads us to
  heuristic predictions about the class of future quasistatic quintessence potentials.
\end{abstract}
\pacs{98.80.-k, 98.80.Cq}
\maketitle
\section{A Low-Density Accelerating Flat Universe}

Observations of high-redshift Type Ia supernovae \cite{snIa} show that
the expanding Universe is today
accelerating, and  was most likely
preceded by a period of deceleration \cite{Turner}. Recent
observations of the Cosmic Microwave Background have shown that
the density of clustered matter is low, $\omm \sim 1/3$, and that
the Universe is flat to high precision, $\Omega_0 = 0.99\pm
0.03$ \cite{CMBR}. Together with the SNIa data, this implies for the
remaining {\em smooth energy} $\omq\sim 2/3$.  The flat
cosmology is therefore determined by one parameter, the present ratio
of smooth to clustered energy,  $\rat\equiv
\omq/\omm = 2.2^{+0.7}_{-0.5} \sim 2$   \cite{mr}, and by the equation of
state $P=P(\rho_Q,a)$, which changes with cosmological scale $a$.

The conventionally-defined deceleration parameter $-a\ddot a/\dot
a^2$ is thus at present
\be
q_0=(1+3w_{Q0}\omq)/2 < 0 \quad
\mbox{or} \quad w_{Q0}<-1/3\omq .
\ee
For $\omq=2/3$, this requires
$-1\leq w_{Q0}< -1/2$. (SNIa data have recently established a
slightly stronger limit $\dub <-0.55$ at $95\%$ confidence \cite{wQ}.)
Within these limits, we will be concerned to distinguish between
static smooth energy $\dub=-1$ and the other limiting case
$w_{Q0}=-0.5$, the maximum value for which the present universe will
be accelerating \cite{dub}.

The observations thus require smooth energy $\rho_Q$ with negative pressure
$P=\dub\rho_Q$. If this is
quantum vacuum energy, its small present value would require fine-tuning to
54 decimals. More plausibly, we assume the
vacuum energy to vanish exactly, due to some unknown symmetry
principle, and invent some dynamical mechanism for anti-gravitating
smooth energy, slowly varying on a Hubble scale.
This explains how a huge cosmological ``constant'' could become small
now, but does not fix the one
cosmological parameter $\rat\sim 2$. Unless totally accidental, the ``cosmic
coincidence'' of  our presence just when the dark energy is dominating the background
energy, must derive either from a still-to-be discovered
quantum cosmology or from some selection effect connected with
our existence (Anthropic Principle).

By quintessence \cite{Quint,SWZ} we mean a rolling scalar field $\phi$
which, for a large class of initial conditions, converges towards a
{\em tracker field} obeying an effective ``equation of state'' (EOS)
$w_Q(a) = P/\rhoq$ with $P\equiv P_Q(\rhoq,a)$,  which tracks the
background field "equation of state" $w_B(a)$. The background EOS
$w_B(a)=P_{rad}/(\rho_{rad}+ \rho_{mat})$ reduces gradually from $1/3$
in the radiation-dominated era to zero in the matter-dominated era,
changing appreciably only since the era of radiation-matter equality,
$1+z_{eq}=1/a_{eq} =3880$. (We are assuming a flat Universe with
$h^2=0.5$  \cite{HST}, so that at present, the critical density $\rho_B+\rhoq$
is $0.940E-29~ {\rm g~cm}^{-3}=4.05E-47~{\rm GeV}^4$,  the
quintessence/matter ratio is $\rat=2$, and the background
density is $\rho_{B0}=1.35E-47~{\rm GeV}^4$.)

In the next three sections, we consider tracking scalar field dynamics
generally, the parametrization of two different dynamical quintessence
models, and other possibilities for the smooth energy.  In Section 5,
we review a recent calculation \cite{cosmo8} showing that, for $\rat
\sim 2$, the asymptotic collapsed mass fraction is maximal so that the
evolution of intelligent observers is most likely.  Finally, this
observation is used to infer anthropically and heuristically a
testable constraint on future theories of quantum cosmology and the
static or quasi-static nature of the smooth energy.

\section{Scalar Field Dynamics: Distinguishing Dynamic from Static
  Smooth Energy}

The equation of motion for the scalar field follows from the
flat-space Friedman equation \be H^2=(8\pi G/3)(\rho_B+\rho_Q),
\ee and two coupled equations for quintessence energy
conservation and for the evolution of $\rhoq(a)$ and $\phi(a)$

\be
-d\rhoq/dN  =  3\dot{\phi}^2 =6(\rhoq-V) =n_Q\rhoq \ee
\be  {{d\phi}\over{dN}}  =  (\dot{\phi}/H)=  \sqrt{6(\rhoq-V)/(8\pi G\rho)} =\sqrt{n_Q\Omega_Q}\cdot M_P .
\ee
On the right-hand side, $\rho=\rhoq+\rho_B,~\rho_Q=\phidot^2/2 +
V,$ and $P=\phidot^2/2-V$.  The two
dependent variables $\rhoq$ and $\phi$ are functions of the
independent variable $N \equiv \ln a$ . Hereafter, we use reduced
Planck units $8\pi G \equiv M_P^{-2}=1$ for $\phi$.

Using $V(\phi)=\rhoq(1-\dub)/2$ and defining \cite{SWZ} a function
\begin{widetext}
\be \Delta(a) \equiv d \ln V/d\ln\rhoq=1+d\ln (1-\dub)/d\ln\rhoq=
1+(d\dub/dN)/3(1-\dub^2) , \ee we can rewrite Eq. (3) as \be
d\dub/dN=3(1-\dub^2)(\Delta -1), ~~\mbox{where} ~~~~ \Delta =
(-V'/V)\sqrt{\Omega_Q/n_Q} . \ee\end{widetext}

Defining  $x\equiv\dot\phi^2/2V(\phi)$, the ratio of
quintessence kinetic and potential energies,
equation (6) shows $\Delta(a)-1=d\ln
x/6\cdot dN$ to be appreciable only
where the quintessence field is changing rapidly. Thus $\dub$ varies
slowly where the field is either nearly frozen
($\dub\sim -1$) or is tracking ($\Delta \sim 1)$.

Starting from a broad range of initial conditions, $w_Q(\phi)$
oscillates between $\sim 1$ and $\sim -1$ and enters an epoch in
which the energy density $\rho_Q(\phi)$ is frozen at a relatively small
value.  Then, this field converges onto a tracker solution in
which the slope $-d \ln \rhoq/dN \equiv n_Q$ is nearly constant
at a value less than $-d \ln \rhob/dN \equiv n_B$ for the
background matter. Here $N\equiv \log a$, and $n_i \equiv
3(1+w_i)=-d \ln \rho_i/d N$ is the adiabatic index: $n_B=4$ and $3$
for radiation and matter, respectively.

\section{Parametrization of Dynamical Quintessence by Two Different
  Equations of State}

The condition for a tracker solution to exist is \cite{SWZ} that
$-V'/V$ be a slowly decreasing function of $\phi$ or $N$, where
$V'\equiv dV/d\phi$, or that \be 0<\Gamma(a)-1 \equiv
d(-V/V')/d\phi \ee be nearly constant. It is convenient to
measure the steepness of the potential by the logarithmic
derivative $\beta(\phi) \equiv -d\ln V/d\ln\phi$, so that
\be\Gamma-1=d(\phi/\beta)/d\phi=1/\beta(\phi)\cdot(1-d\ln\beta/d\ln\phi).\ee
We will consider below, an inverse-power potential for which
$\beta$ is strictly constant, and an ``isothermal'' potential,
for which $\beta$ slowly decreases with $\phi$ so as to keep the
equation of state $\dub=constant$.

The ordinary inflationary parameters are then \be\eta(\phi)
\equiv {{V''}\over{V}} ,~~2\epsilon\equiv ({{V'}\over{V}})^2,
\Gamma={{V\ddot V}\over{\dot
V^2}}={{\eta(\phi)}\over{2\epsilon(\phi)}}.\ee When
$\eta,~2\epsilon\ll 1$, a tracking potential will be {\em
slow-rolling}, meaning that $\ddot\phi$ in the scalar field
equation of motion and $\dot\phi^2$ in the kinetic energy are both
negligible. This slow-roll approximation is usually satisfied in
ordinary inflation.  In the early e-folds of quintessence,
however, $-V'/V=\beta(\phi)/\phi$ is slowly changing but is
itself not small for $\beta \neq 0$.  This establishes the
important distinction between static or quasi-static quintessence
($n_Q(1)\sim 0,~w_{Q0}\sim -1$) and sensibly dynamical
quintessence ($n_Q>1,~\dub>-2/3)$.  At the observationally
allowed upper limit $w_{Q0}=-0.5$, the first few scale e-folds
are {\em
  fast-rolling}. Such dynamical quintessence will become slow-roll
only in the very far future, after many e-folds of the
quintessence field slowly driving $\Delta$ from $1$ to zero and
$\dub$ from its present value to zero.  This means that while the
slow-roll approximation is applicable to ordinary inflation,
dynamical quintessence generally requires exact solution of the equations
of motion (4) and (6).

For tracking potentials derived from physical principles, it is
reasonable to assume that $\beta \approx constant$, at least since
tracking began.  We therefore parametrize quintessence so that the
SUSY-inspired potential \be V(\phi)=V_0(\phi_0/\phi)^\beta,~~~~
\beta=3.5 \ee is inverse power law, and $\Gamma
-1=1/\beta=constant$.

For comparison, we also consider the isothermal equation of state
\be V(\phi)=V_0[\sinh(\alpha\phi_0)/\sinh(\alpha\phi)]^{\beta},\ee
where $\beta=2, \alpha \equiv \sqrt{3/(2\beta+\beta^2)}=0.612 $.
Then $\Gamma -1=\Omega_B/\beta$ is not constant, but decreases
from $1/\beta$ when tracking begins, to zero in the far future.
This potential is called ``isothermal'' because for it, $\dub=
-2/(2+\beta)=-1/2$ and $n_Q=3\beta/(2+\beta)=3/2$ are constant
once matter dominates over radiation. This makes the tracker
approximation exact for the isothermal potential: $\Delta=1$, Eq.
(6) is trivially satisfied, and Eq. (4) makes $-V'/V=n_Q \cdot
da/a$. This potential allows a scaling solution $V(a),~\rho_Q(a)
\sim a^{-n_Q}$ \cite{Matos} and has the inverse power potential
$V=V_0(\phi_0/\phi)^{\beta}$ and the exponential potential
$V=V_0\exp(-\sqrt{n_{Q}} \phi)$ as limits for $\alpha\phi\ll 1$
and $\gg 1$, respectively.  It therefore interpolates between an
inverse power potential at early times and an exponential
potential at late times.

\begin{figure}
\resizebox{8.5cm}{!}{\includegraphics{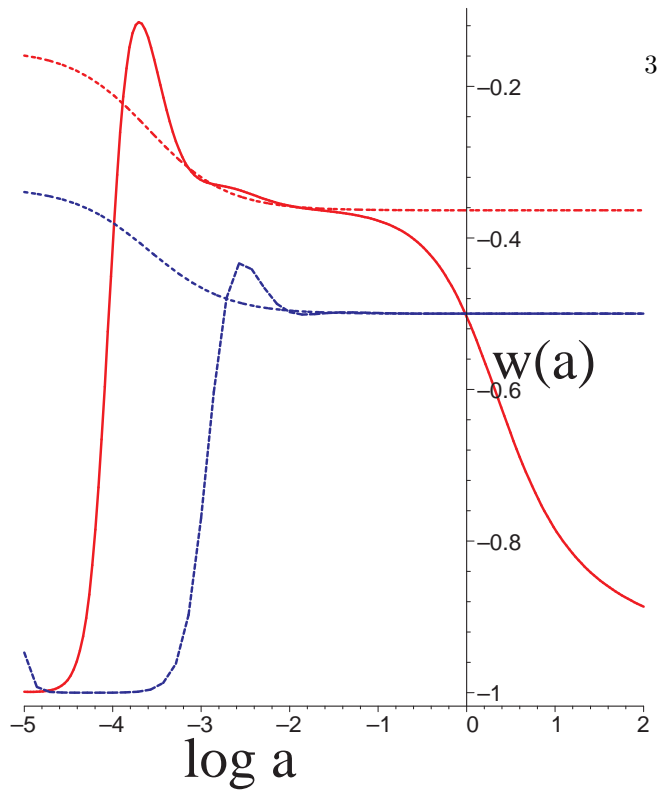}}
  \caption{\label{f1}Exact and tracker approximation solutions $\dub$ for inverse power and
  isothermal potentials chosen to give present
  values $\rat=2,~~w_{Q0}=-0.5$.  The exact $\dub$ values for the
  inverse-power and the isothermal quintessence potentials (11) and (12)
  are shown by the solid and dashed curves. The tracker approximation
  (dash-dotted curves) is exact for the isothermal potential, but holds
  only briefly for the inverse-power potential and considerably
  overestimates $\dub$ once quintessence is appreciable.}
\end{figure}

In each of these potentials, the two parameters have been chosen to
fit the present values $\rat=2$ and
$\rho_{Q0}=2\rho_{cr}/3=2.7E-47~GeV^4$, so that for $w_{Q0}=-0.5$,
$n_{Q0}=3/2$ the present value of the potential is
$V_0=V(\phi_0)=1.013E-47~GeV^4$. This requires tracking to begin
after matter dominance,  reaching present values $\phi_0=2.47$ for
the inverse-power and $1.87$ for the isothermal EOS, respectively.
Tracking continues indefinitely for the isothermal EOS, but is of
relatively short duration for the inverse-power potential (Fig.~\ref{f1}).
We believe these parametrizations of $V(\phi)$ to be representative of
reasonable smooth potentials, over the red-shift range $z<5$
that is observationally accessible. If we considered lower values of
$\beta$ for the present ratio $\rat=2$, $w_{Q0}$ would decrease from
$-1/2$ to $-1$ as the potential became faster rolling.

For the inverse-power potential, $-V'/V= \beta/\phi$, and for the
isothermal potential $-V'/V=\beta\alpha/\tanh(\alpha\phi)$,
respectively, so that, for $w_{Q0}=-0.5$, $\eta$ and perforce
$\epsilon$ are not now small.  Thus both the dynamical quintessence
potentials we are considering are now still fast-rolling.  The
isothermal potential, with $\beta=2$, asymptotically approaches
$~\exp(-1.22~\phi)$ and will never be slow-rolling.  The inverse-power
potential, with $\beta=3.5$, will become slow-rolling when
$\phi>\beta$ and will asymptotically approach a de Sitter solution in
the distant future. In the observable recent past, its $\dub$
increases with $z$ approximately as $\dub(z)\approx-0.5+0.016z$.

\subsection{Tracker Approximation}

After transients depending on initial conditions, and a frozen
epoch ($\dub\approx -1$), the scalar field overshoots and then
converges onto a solution \be
n_Q(a)=n_B(a)\beta/(2+\beta)=n_B(a)/2\ee
which tracks the
background $n_B(a)$ (Fig.~\ref{f1}). During the tracking regime, until
quintessence dominates, $\Delta\approx 1$ and $\Omega_Q\approx
n_Q\phi^2/\beta(\phi)$ increases quadratically with field
strength.

Driven by the background EOS which is changing around $z_{cr}=3880$,
the quintessence equation of state parameter $\dub(a)$
slowly decreases. The inverse-power potential reaches its tracker at
$\log a \approx -3.2$, but remains there only until $\log a \approx
-1.8$, when the growth of the quintessence field slowly drives $\dub$
down from the tracker value $-0.364$ towards $-1$ in the very far
future (solid curve in Fig.~\ref{f1}).  For the inverse-power potential, the
tracker approximation holds only briefly, and thereafter seriously
overestimates $\dub$. The exact isothermal solution reaches
its tracker later, at $\log a >-1$, but remains exactly on tracker thereafter
(dashed curve in Fig.~\ref{f1}).

\begin{figure}
\resizebox{8.5cm}{!}{\includegraphics{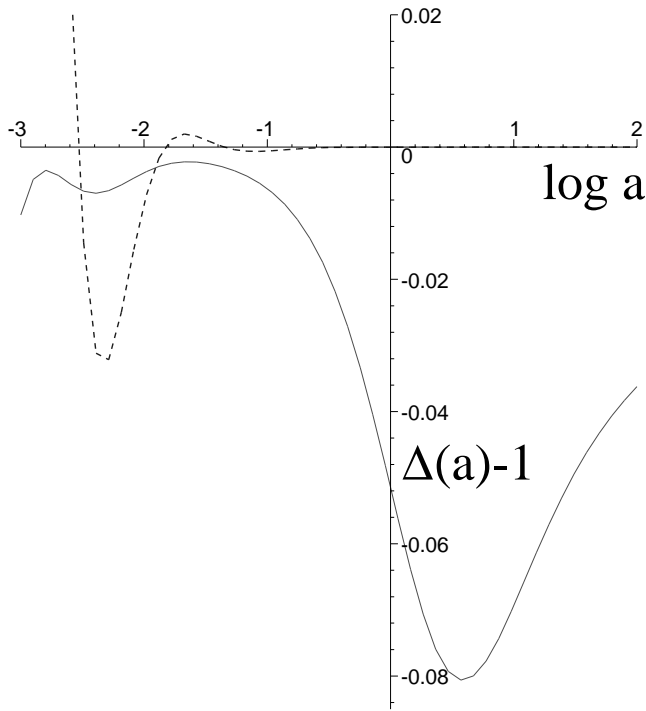}}
  \caption{\label{f2}
The quantity $\Delta-1=d\ln x/6\cdot d\ln a$ for the inverse-power and the isothermal
  quintessence potentials (10) and (11) are shown by the solid and
  dashed curves. $\Delta$ differs from
  unity only before tracking and, in the case of the inverse-power
  potential, where the quintessence field is growing.}
\end{figure}

Fig.~\ref{f2} magnifies the changes in $\dub$ that
transpire before and after tracking, by
plotting $d\ln x/6\cdot dN$, where $x$ is the ratio of
quintessence kinetic to potential energy.  Fig.  3 shows the
integrated effects of $n_Q(a)$, by plotting the integral of equation (3).

Once reaching the tracker, the isothermal  equation of state
stays constant at $\dub(\phi)=-2/(2+\beta)=-1/2$.
The evolution with cosmological scale is fixed by the
scaling $V/V_0=\rho_Q/\rho_{Q0}=a^{-3/2}$, which makes (cf. Fig.~\ref{f4})
\be
\mathcal R(a)=\sinh^2 (\alpha \phi),\quad \Omega_Q=1-\Omega_B=\tanh^2
(\alpha \phi).
\ee

For the present ratio $\rat \sim 2$ one has $\alpha \phi_0\leq
1.146$ and $\phi_0 \leq 1.87$. The observations that the Universe
is now accelerating \cite{dub} fix the bounds
\be
-2/(2+\beta) \leq -1/2,~~~\beta \leq 2, ~~~\alpha\geq
0.612,~~~\phi_0 \leq 1.87.
\ee
For $\beta=0$, the potential is
static: $\Lambda$-dominance started at redshift
$R_0^{1/3}-1=0.260$, the universe was then already accelerating
$-q=0.333$, and is now accelerating faster $-q_0=0.5$.  For
larger $\beta$, the scalar field is more dynamic.  At the upper
limit, $\beta=2$ the isothermal potential rolls only as fast as
the observed acceleration ($-q_0 \geq 0$) allows, tracking starts
only after matter domination, and acceleration starts only now.
Nevertheless, $\rho_Q(a)$ was still subdominant to the radiation
density all the way back into the era of Big Bang nucleosynthesis
\cite{Bean}. In Sec. 5, we will consider these two limits for the
isothermal smooth energy $\beta=2$ (QCDM) and  $\beta=0$ (LCDM).

As mentioned above, the tracking approximation is exact for the
isothermal potential, but only briefly valid for the inverse-power potential.
For $\alpha\phi\ll 1$, the background dominates and the isothermal
equation of state reduces to the inverse power potential $V\sim
\phi^{-\beta}$, for which, along the tracker, $\phi,~\sqrt{\mathcal
  R}\sim a^{3/(2+\beta)}$. We shall, however, also be interested in
the quintessence dominated era $\alpha\phi>1$, when different
potentials give different present and future behavior.

\subsection{Present and Future of the Quintessence-Dominated Universe}

\begin{figure}
\resizebox{8.5cm}{!}{\includegraphics{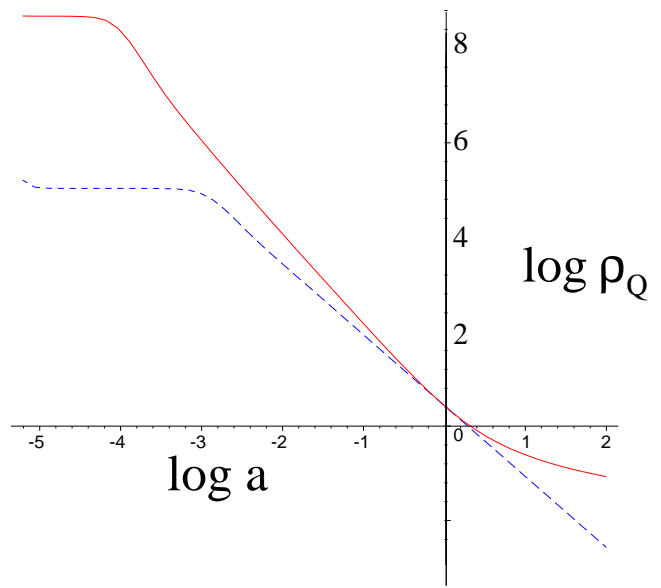}}
  \caption{\label{f3}
Logarithmic plots of the quintessence energy density, in
  units $10^{-47}$ GeV$^4$, for the inverse-power (solid curve) and
   isothermal (dashed curve) potentials.
  After transients, but before tracking, the fields are frozen.  On tracker,
  the adiabatic index in $\rhoq\sim a^{-n_Q(a)}$ stays at 3/2 for the
  isothermal EOS, but continuously decreases towards zero for the
  inverse-power EOS. The present density $\rho_{Q0}=2\rho_{cr0}/3$
  and adiabatic index $n_{Q0}=1.5$ are the same for both.}
\end{figure}
\begin{figure}
\resizebox{8.5cm}{!}{\includegraphics{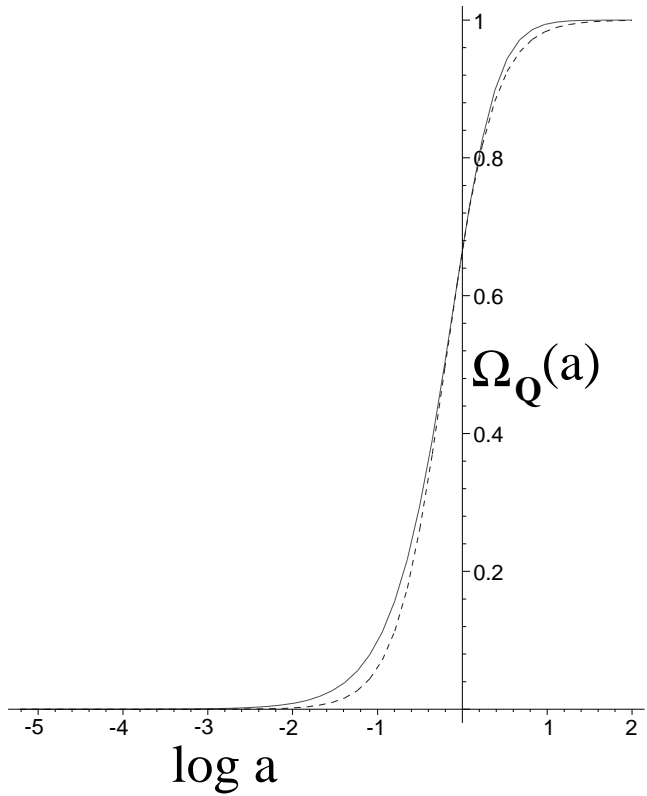}}
  \caption{\label{f4}
Evolution of the quintessence fraction of the energy density.
  $\Omega_Q(a)$ starts growing as $n_Q\phi^2/\beta(\phi)$ at tracking,
 reaches $2/3$ and is changing
  most rapidly at present. It will reach unity only in the distant future.}
\end{figure}

The difference between the two potentials (10,11) appears only in the
evolution of $\dub$ now and in the future, when $\alpha\phi>1$.  For
the inverse power potential, the growth of quintessence drives
$\Delta(a)$ below unity and $w_Q(a)$ decreases slowly towards -1. The
quintessence energy density $\rhoq$ and the expansion rate
$H\rightarrow constant$, drive ultimately towards a de Sitter
universe, which inflates exponentially.  The isothermal EOS will
approach the exponential potential, $\dub$ stays constant, $\rhoq$ and
$H$ continue to decrease, and inflation is power-law.  For both these
potentials, acceleration continues indefinitely, so that two observers
separated by fixed coordinate distance, ultimately have relative speed
$\geq c$.  Because an event horizon exists in both cases, a local observer
cannot construct an S-matrix.  This shows that neither of our
phenomenological potentials is derivable from field theory or string
theory. To be string-inspired and lead to an event horizon, the
quintessence EOS would have to ultimately change back from
accelerating to decelerating ($\dub>-1/3$).

Fig.~\ref{f4} shows the
evolution of the quintessence fraction of
the total energy density $\Omega_Q(a)$ from $0$ to $1$, presently
passing through $\Omega_{Q0}=2/3$.  This evolution is similar for
both potentials, although the inverse-power potential is more
dynamic than the isothermal EOS.

\section{Other Possibilities for Smooth Energy}

The inverse-power potential (10) is a reasonable phenomenological
description for potentials which are not decreasing too fast at
present. The isothermal potential $\dub= constant$ is as good as any other
potential we might choose phenomenologically in the near future, and it
allows an analytic solution.

If $\dub$ is not constant, we could still reconstruct the quintessence
potential from $\dub(z)$.  Indeed $d \dub(z)/dz$ is generically positive for
quintessence and generally negative kor k-essence \cite{Picon},
an alternative in which the scalah field has a non-linear kinetic
energy instead of a potential.  In principle, $\dub(z)$ is observable
in high red-shift supernovae \cite{snIa}, in cluster evolution \cite{Newman,Haiman}
and in gravitational lensing \cite{Cooray}.  In
practice, $\dub(l)$ is poorly constraived observationally \cite{wQ}
and theorbtically \cite{Barger,Limitations,Astier}. Over the small
redshift range for which smooth energy dominates and for which
sensitive measurements are possible, the effects of varying $\dub(z)$
are much smaller than the present uncertainties in measurement and in
cosmological model.  Theoretically, the luminosity distance to be
measured in high-redshift supernova and in comoving volume number
density measurements depends on integrals which smooth out the
sensitivity to $\dub(z)$ \cite{Limitations}.  Until $\omm$ is
determined to (3-2)\% accuracy, SNAP \cite{SNAP} and other future
experiments will only be able to determine some $\dub$ effective over
the observable redshift range $z<2$, to tell whether the smooth
energy is static or dynamic.  If we wait long enough,
the differences among different quintessence potentials will
asymptotically become substantial.

In the next section, we will study how different structure evolves to the present
$\rat=2$  for $\dub=-1$ (LCDM) and for $\dub=-1/2$ (QCDM).
We should interpret this constant $\dub=-1/2$
chosen  as come maximal average value of $w(z)$.  The present
maximal value $w_{Q0}$ would then be somewhat less than -1/2 for quintessence
and somewhat greater for k-essence.

\section{Evolution of Large Scale Structure}

Since structure evolves only in a matter-dominated universe, its
evolution started about the time of radiation-matter equality and
ended about matter-quintessence equality $\mathcal R=1$ or at
scale $a=\rat^{1/3\dub}$ i.e. at redshift
$\rat^{1/3}-1=0.260,~\rat^{2/3}-1=0.53$ for LCDM and for our
isothermal $\dub=-1/2$ QCDM respectively.

Using an improved Press-Schecter formalism \cite{cosmo8}, holding
fixed all other dimensionless constants, we calculated as
function of $\rat$ the fraction of mass that would have already
collapsed into structures larger than $1-2$ Mpc. We interpret this
asymptotic collapsed mass fraction as a measure of the relative
likelihowd of large-scale structures, galaxies, habitable solar
systems and intelligent life. (There is a tacit assumption here that
intelligent life depends critically on the evolution of large scale
structure and is therefore relatively rare in our universe.)  Not
surprisingly \cite{Wein}, the observed $\rat \sim 2$ fell within the
range of small cosmological constants or smooth energy density for
which intelligent life is likely.  Surprisingly, this likelihood was
{\em maximal} for our QCDM and nearly maximal for LCDM

\section{The Anthropic as a Heuristic Principle}

Quintessence explains how the cosmological ``constant'' $\Omega_Q$ could
become small, but not why it is just now comparable to $\Omega_B$ and
the Universe is now accelerating.  Hopefully, the one remaining
cosmological parameter $\rat\sim 2$ will someday be derived from
fundamental theory.  Meanwhile, in the face of what is now known, what
is the least biased inference to be made from the observed datum $\dub\sim
2$, at or near the peak of the logarithmic asymptotic mass
distribution?

Applying probability considerations to a unique
universe is problematic.  Nevertheless, we have now become
accustomed to the role of observers in quantum mechanics.  This
encourages us, at least tentatively, towards an anthropic explanation
of the cosmic coincidence, why we live at a time when the smooth
and clustered energy densities are approximately equal.  Following
Vilenkin's {\em Principle of Mediocrity} [22], we assume that, out of a
{\em conceivable} ensemble of flat cosmologies, our universe or
cosmology is typical of those permitting intelligent life.  Then the
observed datum suggests that the entire prior distribution in $\rat$
must be flat, at least for $\rat=\mathcal O(1)$.  But, if we are
decided to apply probability to the Universe,  over what is the
prior a distribution?

Proponents of the Anthropic Principle
\cite{Wein,AP} have assumed a prior
distribution of real universes, with different values of $\rat$,
almost all of which are inhospitable to life.  Since these many other
universes are practically unobservable, this application has no
practicable predictive value, and is rightfully criticized as
unscientific. It is important, however,  to remember that the prior is a
functional of (subjective) {\em hypotheses}, not of the data sample.

Instead of such a prior distribution of {\em universes}, we
\cite{cosmo8} assumed a flat prior probability distribution,
\be
\mathcal P_{*}  \propto 1/(-V'(\phi)) \qquad \mbox{slowly-varying},
\ee
for (future) fundamental {\em theories}.  This does not require that
the potential be slow-rolling ($\eta,~\epsilon <<1$). If prospective
measurements show $w_{Q0} \approx -1$, then
Eq. (15) requires a quintessence
potential of generic form like $V(\phi)=V_1 f(\lambda\phi)$, where
$V_1 \sim \mathcal O(M_P^4)$ is a large energy density, $f(x)$ is a
dimensionless function involving no very large or very small
parameters, and $\lambda$ is a very small dimensional parameter
\cite{Wein0}.   Because such a prediction for
the static or quasistatic quintessence potential is theoretically and observationally
falsifiable, our way of applying anthropic reasoning to theories is
scientific, while the application to subuniverses is not, at least not
until these other universes can be observed.  This application of
anthropic reasoning is only heuristic, as was the Equivalence
Principle which guided us towards General Relativity and the
Correspondence Principle which guided us towards quantum mechanics.

\begin{acknowledgments}
We thank Philip Moll for helping edit the figures.
\end{acknowledgments}

\end{document}